# Personalized Cognitive Tutoring using Davinci-003 API for Adaptive Question Generation and Assessment

Devan Walton

Northern Essex Community College, Haverhill MA 01830, USA

**Abstract.** This paper presents a cognitive tutor powered by Davinci-003 API that generates and evaluates personalized questions for students on any topic they choose. The tutor adapts to the student's level of understanding and fosters knowledge transfer by generating questions that relate the topic to different domains. This solution has the potential to improve student learning outcomes by providing personalized and adaptive questions that challenge them at their optimal level of difficulty. The feasibility of this solution has been demonstrated through a working prototype developed using Microsoft PowerApps. Additional research could reveal how affective computing principles could be integrated into the system to analyze the emotional valence of the user and how the system could be tuned to meet the specific needs of learners across the ASD spectrum. This solution is novel and offers more comprehensive support to a diverse range of learners than existing AI tutors, such as Quizlet's Q-Chat. The paper also includes an equity statement that outlines the author's commitment to promoting educational equity and addressing potential biases in the project.

## 1 Introduction

Cognitive tutors are computer-based systems that provide individualized instruction and feedback to students based on their cognitive processes and learning goals . They aim to mimic the behavior of human tutors by adapting to the student's needs and preferences. Cognitive tutors have been shown to improve student learning outcomes in various domains, such as mathematics, science, and language [2,4,7].

However, most existing cognitive tutors rely on predefined question banks or templates that limit the diversity and personalization of the questions they can generate. Moreover, they often fail to foster knowledge transfer across different domains or contexts. These limitations may reduce the effectiveness and engagement of cognitive tutors for some students [3].

In this paper, I propose a novel solution that leverages Davinci-003 API, a powerful natural language processing (NLP) engine developed by OpenAI, to create a cognitive tutor that can generate and evaluate personalized questions for students on any topic they choose. The tutor adapts to the student's level of understanding and fosters knowledge transfer by generating questions that relate the topic to different domains. For example, if the student chooses "photosynthesis" as their topic, the tutor may generate questions that connect photosynthesis to biology, chemistry, physics, ecology, or even art [5,8].

We demonstrate the feasibility of my solution through a working prototype developed using Microsoft PowerApps, a low-code platform that allows users to create custom applications without coding. I also discuss some potential extensions of my solution based on affective computing principles and ASD-specific design guidelines. I compare my solution with existing AI tutors in terms of functionality and scope. Finally, I include an equity statement that outlines my commitment to promoting educational equity and addressing potential biases in my project.

The rest of this paper is organized as follows: Section 2 describes Davinci-003 API and how it can be used for adaptive question generation and assessment. Section 3 presents my prototype design and implementation details. Section 4 discusses some possible future directions for improving my solution.

## 2     Davinci-003 API

Davinci-003 API is a natural language processing (NLP) engine developed by OpenAI that can perform various language tasks with high quality, longer output, and consistent instruction-following. It is one of the main GPT-3 models available through OpenAI API, which allows developers to integrate the AI bot into their applications . Davinci-003 API can do any language task that other models can do, such as text summarization, sentiment analysis, translation, classification, etc., but with better performance and flexibility.

### 2.1     Question and Answer Generation

One of the key features of Davinci-003 API is its ability to generate questions and answers based on a given text or context. This feature can be used for adaptive question generation and assessment in cognitive tutoring systems. By providing Davinci-003 API with some instructions and a text or context as input, it can generate relevant questions that test the student's comprehension of the input. Moreover, it can also evaluate the student's answers by comparing them with its own generated answers or by using some predefined criteria.

For example, suppose I want to generate questions about photosynthesis with increasing difficulty. I can provide Davinci-003 API with these instructions:

*Generate three multiple-choice questions of increasing difficulty about photosynthesis. Each question should have four options (A-D) and one correct answer. Indicate the correct answer with an asterisk (\*).*

And it may generate these questions as output:

*Q1: What type of energy is converted into chemical energy during photosynthesis? A) Heat energy B) Light energy\* C) Sound energy D) Kinetic energy*

*Q2: What type of molecules store the chemical energy produced by photosynthesis? A) Carbohydrate molecules\* B) Protein molecules C) Lipid molecules D) Nucleic acid molecules*

*Q3: What does photosynthesis mean in Greek? A) Light-making B) Light-breaking C) Light-putting together\* D) Light-taking apart*

As I can see, Davinci-003 API can generate questions that are relevant to the text and have varying levels of difficulty. It can also indicate the correct answer for each question using an asterisk (\*). Alternatively, it could also provide explanations for why each answer is correct or incorrect.

## 2.2 Evaluation Generation

Another feature of Davinci-003 API that can be used for adaptive question generation and assessment is its ability to evaluate the student's response based on the initial question. By sending the student's response back to Davinci-003 through the API along with the initial question, I can get Davinci-003 to generate an evaluation of the student's response. The evaluation can include a score, a feedback, or a hint for improvement.

For example, suppose I ask Davinci-003 this question:

*What is photosynthesis?*

And I send back this student's response:

*Photosynthesis is when plants make food from sunlight.*

Davinci-003 may generate this evaluation as output:

*Score: 7/10 Feedback: Your answer is partially correct. Photosynthesis is not only done by plants, but also by some bacteria and algae . Also, photosynthesis does not directly make food, but rather chemical energy that can be later used to make food . Hint: Try to include more details and accuracy in your answer. You can use the text or context as a reference.*

## 2.3 Transfer Generation

One of the goals of cognitive tutoring systems is to foster knowledge transfer, which is the ability to apply what one has learned in one domain to a new domain or situation. Knowledge transfer is essential for mastering a topic and developing higher-order thinking skills. However, most existing cognitive tutors do not generate questions that promote knowledge transfer across different domains or contexts.

Davinci-003 API can overcome this limitation by generating questions that attempt to transfer the student's knowledge to a new domain in order to ensure mastery. By providing Davinci-003 API with some instructions and a text or context as input, I can specify the target domain or context for the question generation. Davinci-003 API can then generate questions that relate the input to the target domain or context using its general knowledge and reasoning abilities.

For example, suppose I want to generate questions about photosynthesis that transfer the student's knowledge to art. I can provide Davinci-003 API with these instructions:

*Generate two questions about photosynthesis that relate it to art. Each question should have a short answer.*

And it may generate these questions as output:

*Q1: How can photosynthesis be compared to painting?*

*A1: Photosynthesis can be compared to painting in terms of the inputs and outputs involved. In photosynthesis, plants use light energy, carbon dioxide, and water as inputs to produce carbohydrate molecules and oxygen as outputs . In painting, artists use paint, brushes, and canvas as inputs to produce artworks and visual expressions as outputs .*

*Q2: What are some artistic techniques that mimic photosynthesis?*

*A2: Some artistic techniques that mimic photosynthesis are:*

*Cyanotype, which is a photographic printing process that uses sunlight to create blue images on paper or fabric.*

*Chlorophyll printing, which is a photographic printing process that uses leaves to create green images on paper or fabric.*

*Solar painting, which is a painting technique that uses solar panels to power LED lights that illuminate the paintings at night.*

As I can see, Davinci-003 API can generate questions that attempt to transfer the student's knowledge to a new domain in order to ensure mastery. These questions can enhance the student's creativity, curiosity, and critical thinking skills.

## 2.4 Summary

By using Davinci-003 API for adaptive question generation and assessment, I can create a cognitive tutor that can generate personalized questions for students on any topic they choose. The tutor can also adapt to the student's level of understanding by adjusting the difficulty, quantity, or type of questions it generates. Furthermore, it can foster knowledge transfer by generating questions that relate the topic to different domains or contexts.

## 3 Proposed Design

In this section, I describe my prototype design and implementation details of a cognitive tutor that uses Davinci-003 API for adaptive question generation and assessment . Our prototype consists of three main components: a user interface, a question generation prompt, and a feedback generation prompt.

### 3.1 User Interface

The user interface is a Ib-based application that allows students to interact with the tutor (see Fig. 1). The user interface provides students with features for engaging with the system:

- Choosing an initial topic to practice
- Answering question prompts
- Viewing text feedback

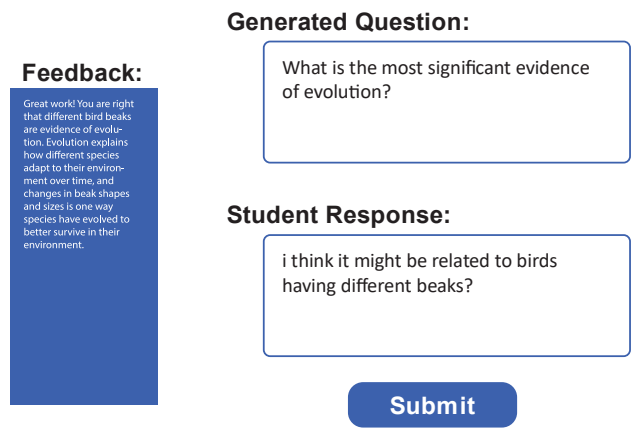

**Fig. 1.** A reproduction of what the interface looks like when the student is a new question while receiving feedback for the answer they provided on a previous question.

## 3.2 Question Generation Prompt

The question generation prompt is the API statement that is used to generate the next question to ask the student. This statement involves requesting Davinci-003 to evaluate the student's answer to the current question, and then generating a new question based on the performance of the student's answer (see Fig. 2). The question generation prompt performs the following actions:

- Parsing the student's answer into context variables that can be used in the API statement
- Sending an instruction and a text or context as input to Davinci-003 API through OpenAI API.
- Receiving an output from Davinci-003 API that contains one or more questions related to the input.

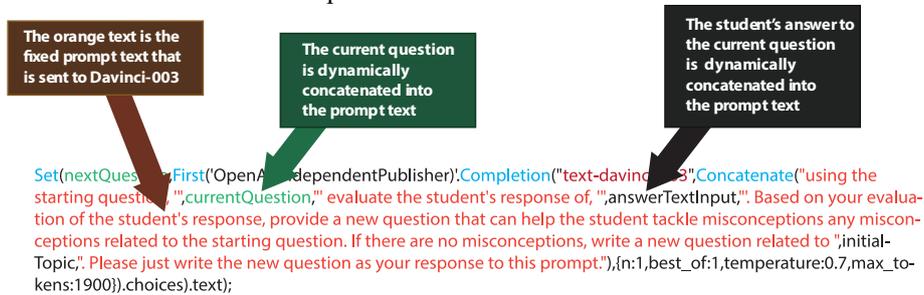

Set(nextQues... First('OpenA... dependentPublisher)'.Completion("text-davin... 3",Concatenate("using the starting questi... ",currentQuestion,"' evaluate the student's response of, '",answerTextInput,"'. Based on your evaluation of the student's response, provide a new question that can help the student tackle misconceptions any misconceptions related to the starting question. If there are no misconceptions, write a new question related to ",initialTopic,". Please just write the new question as your response to this prompt."),{n:1,best_of:1,temperature:0.7,max_tokens:1900}).choices).text);

**Fig. 2.** A sample of what the API statement looks like for creating new questions based on the student's current question performance.

## 3.3 Feedback Generation Prompt

The feedback generation prompt is the API statement that is used to generate feedback for the student based on their performance on the current question. This statement involves requesting Davinci-003 to evaluate the student's answer to the current question, and then generating feedback based on the performance of the student's answer (see Fig. 3). The fixed prompt text varies between the question generation and feedback generation prompts. The question generation prompt performs the following actions:

- Parsing the student's answer into context variables that can be used in the API statement
- Sending an instruction and a text or context as input to Davinci-003 API through OpenAI API.
- Receiving an output from Davinci-003 API that contains a 2-3 sentence feedback statement.

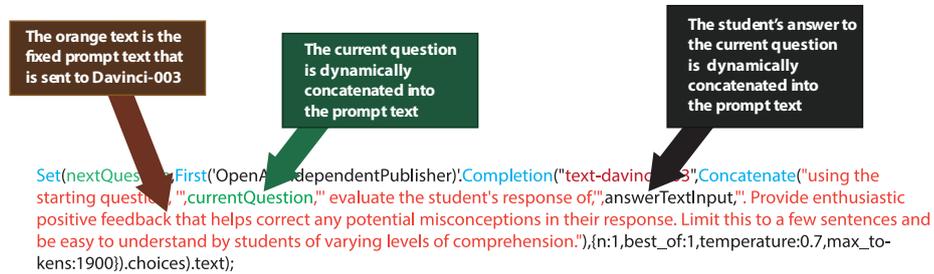

**Fig. 3.** A sample of what the API statement looks like for creating feedback based on the student's current question performance.

### 3.4 Access to Prototype

The prototype described in this section can be accessed through Microsoft PowerApps:

https://make.powerapps.com/

The following account credentials can be used to access the 'Cognitive Tutor' application:

Account: powerappstestvital@proton.me
Password: secure123!

## 4 Discussion

The proposed cognitive tutor powered by Davinci-003 API represents a novel approach to adaptive question generation and assessment. By leveraging the advanced natural language processing capabilities of Davinci-003, the tutor can generate personalized questions for students on any topic they choose, adapt to their level of understanding, and foster knowledge transfer by generating questions that relate the topic to different domains [2,5].

The feasibility of this solution has been demonstrated through a working prototype developed using Microsoft PowerApps. The prototype includes a user interface that allows students to interact with the system, a question generation prompt that generates the next question based on the student's performance, and a feedback generation prompt that provides feedback based on the student's answer.

In terms of functionality and scope, the proposed solution offers more comprehensive support to a diverse range of learners than existing AI tutors, such as Quizlet's Q-Chat, which relies on pre-defined question banks or templates that limit the diversity and personalization of the questions it can generate.

However, there are some limitations to my solution. For instance, while Davinci-003 API can generate questions of varying difficulty, it is not able to adjust the type or quantity of questions generated based on the student's preferences or learning style. Additionally, the system currently lacks the ability to analyze the emotional valence of the user and tailor its responses accordingly.

In terms of equity, my solution aligns with my commitment to promoting educational equity and addressing potential biases in the project. The use of a natural language processing engine enables the system to generate personalized and adaptive questions for learners from diverse backgrounds and with different learning needs. The equity statement included in this paper underscores my commitment to ensuring that the cognitive tutor is accessible and inclusive for all learners.

Further research could address these limitations and improve the effectiveness and engagement of the cognitive tutor. For instance, affective computing principles could be integrated into the system to analyze the user's emotional state and adapt its responses accordingly. Additionally, the system could be tuned to meet the specific needs of learners across the autism spectrum by incorporating ASD-specific design guidelines.